\documentclass[preprint,showpacs,preprintnumbers,amssymb,superscriptaddress,aps,prd,nofootinbib,11pt]{revtex4-1}
\usepackage{graphicx, epsfig, amssymb} 
\usepackage{amsmath, amsfonts}
\usepackage{bm} 
\usepackage{enumitem}
\usepackage[usenames]{color}
\definecolor{navyblue}{rgb}{0.0, 0.0, 0.5}
\usepackage[linktocpage,colorlinks=true,allcolors=navyblue]{hyperref}
\usepackage[caption=false]{subfig}
\usepackage[utf8]{inputenc}
\usepackage{tensor}
\usepackage{soul}
\usepackage{pict2e}
\usepackage{diagbox}
\usepackage{empheq}

\usepackage{float}
\usepackage{comment}
\usepackage{array}
\usepackage{units}
\usepackage{wasysym}

\DeclareMathAlphabet{\mathpzc}{OT1}{pzc}{m}{it}
\newcommand{\ff}{\hat{f}}

\def\l{\left}
\def\r{\right}
\def\p{\partial}

\newcommand{\angarg}{(\theta,\phi)}
\newcommand{\dth}{\frac{d}{d \theta}}
\newcommand{\sh}{ {}_{-s} S_{\ell m}^{a \omega}}

\newcommand{\syjm}{  {}_{-s}Y_{jm} }
\newcommand{\sylm}{  {}_{-s} Y_{\ell m} }
\newcommand{\sumjm}{\sum_{j=s}^{\infty} \sum_{m=-j}^{j}}
\newcommand{\sumjms}{\sum_{j,m}}
\newcommand{\cS}{\mathcal{S}}
\newcommand{\lmws}{{\ell m \omega s}}
\newcommand{\nn}{\nonumber}

\begin{document}

\title{Series reduction method for scattering of planar waves 
\\ by Kerr-Newman black holes}

\author{Tom Stratton}\email{tstratton1@sheffield.ac.uk}
\affiliation{Consortium for Fundamental Physics,  School of Mathematics and Statistics,
University of Sheffield, Hicks Building, Hounsfield Road, Sheffield S3 7RH, United Kingdom \looseness=-1}
\author{Luiz C. S. Leite}\email{luiz.leite@ifpa.edu.br}
\affiliation{Faculdade de F\'{\i}sica, Universidade 
Federal do Par\'a, 66075-110, Bel\'em, Par\'a, Brazil}
\affiliation{Campus Altamira, Instituto Federal do Par\'a, 68377-630, Altamira, Par\'a, Brazil.}
\author{Sam R. Dolan}\email{s.dolan@sheffield.ac.uk}
\affiliation{Consortium for Fundamental Physics,  School of Mathematics and Statistics,
University of Sheffield, Hicks Building, Hounsfield Road, Sheffield S3 7RH, United Kingdom \looseness=-1}
\author{Lu\'{\i}s C. B. Crispino}\email{crispino@ufpa.br}
\affiliation{Faculdade de F\'{\i}sica, Universidade 
Federal do Par\'a, 66075-110, Bel\'em, Par\'a, Brazil}
\begin{abstract}
We present a practical method for evaluating the scattering amplitude $f_s(\theta,\phi)$ that arises in the context of the scattering of scalar, electromagnetic and gravitational planar waves by a rotating black hole. The partial-wave representation of $f_s$ is a divergent series, but $f_s$ itself diverges only at a single point on the sphere. Here we show that $f_s$ can be expressed as the product of a reduced series and a pre-factor that diverges only at this point. The coefficients of the reduced series are found iteratively as linear combinations of those in the original series, and the reduced series is shown to have amenable convergence properties. This series-reduction method has its origins in an approach originally used in electron scattering calculations in the 1950s, which we have extended to the axisymmetric context for all bosonic fields.
\end{abstract}

\date{\today}

\maketitle

\section{Introduction}

The scattering of fundamental fields by the strongly-curved spacetime of a black hole (BH) is of foundational interest. The topic of  time-independent scattering has been studied in detail since the 1960s \cite{Hildreth1964PhDT64, Matzner:1968, Vishveshwara:1970}, and now there exists a substantial literature \cite{Mashhoon:1973zz,Chrzanowski:1976jb,DeLogi:1977dp,Sanchez:1977vz,MatznerRyan1978,Handler:1980un,Matzner:1985rjn,Futterman:1988ni,Andersson:1995vi,Glampedakis:2001cx,Dolan:2006vj,Dolan:2007ut,Dolan:2008kf,Crispino:2009xt,Cotaescu:2014jca,Crispino:2015gua,Sorge:2015yoa,Gussmann:2016mkp,Leite:2017zyb,Nambu:2019sqn, Folacci:2019vtt,Leite:2019,Leite:2019eis,Folacci:2019cmc}. Nevertheless, as yet there are no accurate calculations of scattering amplitudes for electromagnetic ($s=1$) or gravitational ($s=2$) waves impinging on a rotating BH at an arbitrary angle of incidence $\gamma$ (though see Ref.~\cite{Glampedakis:2001cx} for the scalar-field $s=0$ case). A key obstacle to progress is the lack of convergence of the partial-wave series representation of the scattering amplitude $f_s(\theta,\phi)$. In this work, we show that this obstacle may be overcome by applying a series-reduction technique with its roots in the 1950s \cite{Yennie:1954zz}. This work clears the way for accurate numerical calculations of scattering amplitudes in a work to follow.

\begin{figure}
\centering
  \includegraphics[width=0.3\textwidth]{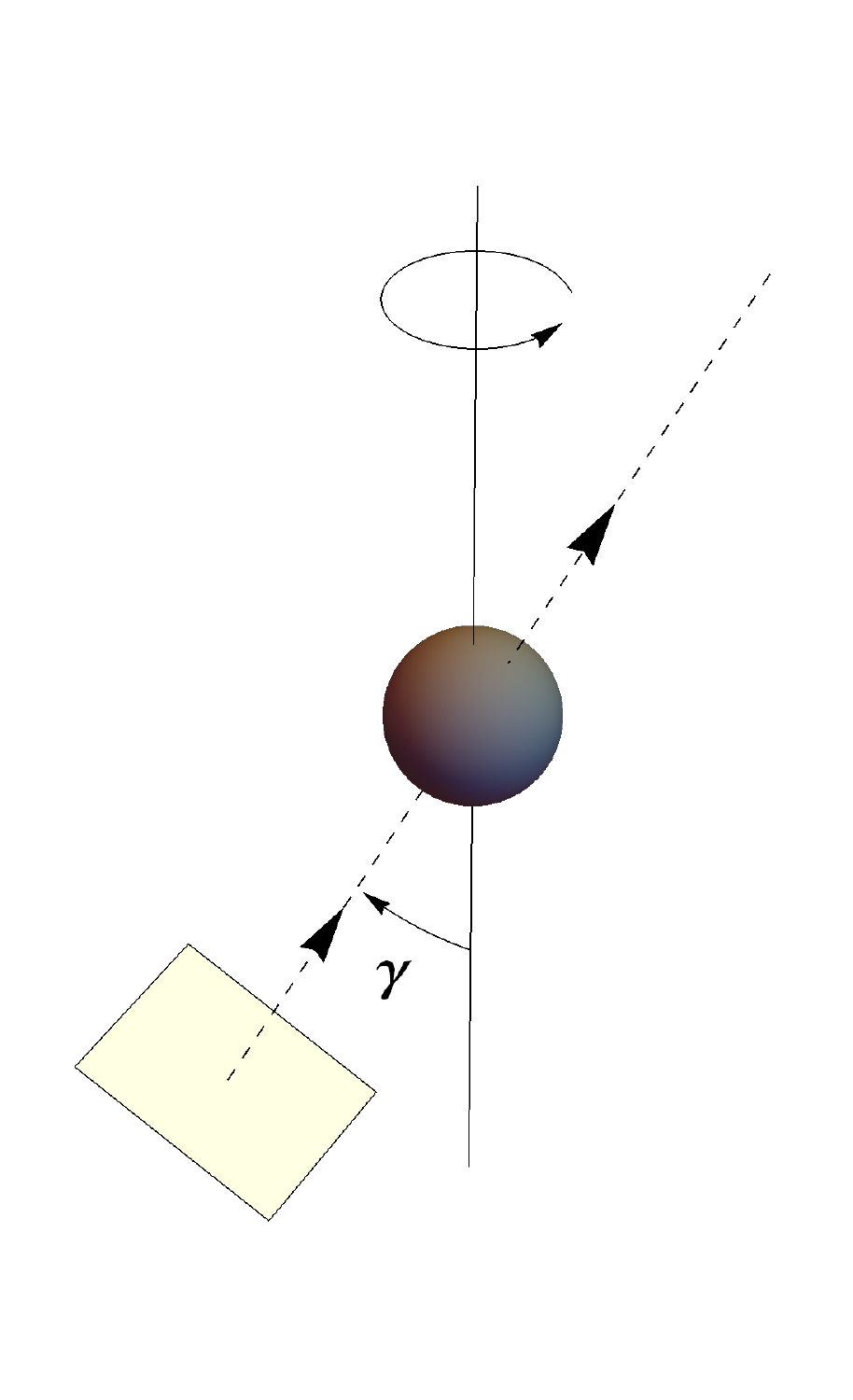}
\caption{In this setup, a planar wave impinges upon a rotating black hole in the direction specified by the angles $\theta_0=\gamma, \phi_0=\pi/2$.}
\label{fig:geometry}
\end{figure}

The scenario we consider here is that of a monochromatic planar wave propagating in vacuum, of spin $s$ and circular frequency $\omega$, which impinges upon a gravitating body of mass $M$, such that $\gamma$ is the angle between the direction of incidence and the symmetry/rotation axis of the body (see Fig.~\ref{fig:geometry}). The gravitational field is long-ranged, with a Newtonian-type  $1/r$ potential in the far-field. The long-range nature of the field has three key effects. First, far from the object ($r \gg r_g$ with $r_g \equiv G M / c^2$), the planar wavefronts are distorted by a logarithmic phase term. Second, regardless of the composition of the body, rays in the weak-field ($r \gg r_g$) are deflected through an angle $\theta$ which is inversely proportional to the impact parameter $b$ (cf.~the Einstein deflection angle). Third, due to scattering in the weak field, the scattering amplitude $f_s$ has a physical divergence in the forward direction, that is, at the point on the sphere which is antipodal to the incident direction. A consequence of the physical divergence in $f_s$ is that its representation as an infinite sum over partial waves is not convergent. This is the issue we address herein.

In the scalar field case ($s=0$), Glampedakis and Andersson \cite{Glampedakis:2001cx} overcame the convergence issue by splitting the amplitude $f_0$ into a `Newtonian' amplitude $f_0^{(N)}$ and diffraction amplitude $f_0^{(D)}$, with the former encapsulating the divergence due to the long-ranged nature of the field, and the latter the main diffraction effects arising from the lower-$l$ partial waves. The Newtonian amplitude was written in closed form and shown to diverge at the expected angle, and the diffraction amplitude was calculated from a mode sum with amenable convergence properties. In principle, this method could be extended to higher spin $s$, but here we prefer to develop an alternative method based on that introduced in 1954 in Ref.~\cite{Yennie:1954zz}, and first applied in the BH context in Ref.~\cite{Dolan:2006vj} (see also Ref.~\cite{Dolan:2008kf}), known as the \emph{series reduction method}. 

The rest of the paper is arranged as follows. In Sec.~\ref{sec:partialwave} we review the theory of time-independent scattering in the axisymmetric case. In Sec.~\ref{sec:series-reduction} we present the series reduction method. After reviewing the lack of convergence in the partial-wave series (
\ref{subsec:convergence}) and its physical origin (\ref{subsec:physical}), we introduce the key idea (\ref{subsec:series-reduction}), describe how to decompose the spheroidal harmonics into their spherical counterparts (\ref{subsec:spheroidal}), and how to regulate the series in principle (\ref{subsec:regulation}). In Sec.~\ref{subsec:s0}, \ref{subsec:s1} and \ref{subsec:s2} we obtain the key formulae for the scalar field ($s=0$), electromagnetic field ($s=1$) and gravitational wave cases ($s=2$), respectively. In Sec.~\ref{sec:results} we examine the results of applying the series reduction method in practice, on the convergence of the series (\ref{subsec:error}) and on computing scalar field $s=0$ cross sections (\ref{subsec:csec}). We conclude with a discussion in Sec.~\ref{sec:conclusions}.

\section{Scattering amplitudes and cross sections\label{sec:partialwave}}

The differential scattering cross section for a spin-$s$ wave incident on a Kerr BH can be expressed as \cite{Futterman:1988ni}
\begin{equation}
\frac{d\sigma_s}{d\Omega} = |f_s \angarg|^2 + |g_s \angarg |^2,
\end{equation}
where the helicity-conserving and helicity-reversing amplitudes, $f_s$ and $g_s$, respectively, are given by the partial wave series \cite{Futterman:1988ni}
%
\begin{align} 
\label{eq:amp_fs}
f_s\angarg & \equiv  \frac{ \pi}{i \omega} \sum_{P=\pm1} \sum_{\ell=s}^{\infty} \sum_{m=-\ell}^{\ell}  {}_{-s} S_{\ell m}^{a \omega} (\gamma) \,  _{-s} S_{\ell m}^{a \omega} (\theta) e^{i m (\phi - \phi_0)}  \left( \cS^P_{\lmws} - 1 \right)\hat{\delta}_{s} ,  \\
\label{eq:amp_g2}
g_s\angarg & \equiv \frac{\pi}{i \omega} \sum_{P=\pm1} \sum_{\ell=s}^{\infty} \sum_{m=-\ell}^{\ell} \,  _{-s} S_{\ell m}^{a \omega} (\gamma) \,  _{-s} S_{\ell m}^{a \omega} (\pi - \theta)  e^{i m (\phi-\phi_0)}  P (-1)^{l+m+2} \left( \cS^P_{\lmws} - 1 \right) \, ,
\end{align}
where $\hat{\delta}_{s}=2$ for $s=0$, $1$; and $\hat{\delta}_{s}=1$ for $s=2$. Here $_{-s}S_{\ell m}^{a \omega} (\theta)$ is a spin-weighted spheroidal harmonic (see below), $\cS_{\lmws}^{\pm}$ is the scattering coefficient \cite{Glampedakis:2001cx,Frolov:1998wf}
\begin{align}
\cS_{\lmws}^{\pm} \equiv e^{2 i \delta^\pm_{\lmws}} ,  \label{eq:scat-coef}
\end{align}
and $\delta^\pm_{\lmws}$ is the phase shift, determined from a radial equation. In the case $s = 0$, there is no odd-parity part, and in the case $s = 1$ the phase shift is independent of the parity $P$, and thus the helicity-reversing amplitudes $g_0$ and $g_1$ are identically zero; note the sum over parity in Eq.~(\ref{eq:amp_g2}). In the gravitational-wave case $s=2$, the phase depends on parity according to
\begin{equation}
\frac{\cS_{\lmws}^{+} }{\cS_{\lmws}^{-} } = \frac{\text{Re} \, \mathcal{C} + 12 i M \omega}{\text{Re} \, \mathcal{C} - 12 i M \omega} ,
\end{equation}
where $\mathcal{C}$ is the Teukolsky-Starobinskii constant. Consequently, the helicity-reversing amplitude $g_2$ is non-zero.
The partial wave sum (\ref{eq:amp_g2}) for $g_2$ is convergent, as $\text{Re} \, \mathcal{C} = O(\ell^4)$ in the large-$\ell$ limit. By contrast, the partial wave sum for $f_s$ is not convergent. In the next section we show this, and present a practical remedy.

%
%

%


\section{The series reduction method\label{sec:series-reduction}}

\subsection{Series convergence\label{subsec:convergence}}
The lack of convergence of the partial wave series is most straightforward to demonstrate in the base case of a scalar wave on Schwarzschild spacetime ($s=0,a=0$), for which the scattering amplitude $f_0$ has the representation
\begin{equation}
f_0(\theta) = \frac{1}{2 i \omega} \sum_{\ell = 0}^{\infty} (2 l + 1) \left( e^{2 i \delta_l} - 1 \right) P_\ell (\cos \theta) , \label{eq:f0schw}
\end{equation}
where $P_\ell(\cdot)$ is a Legendre polynomial. In the large-$\ell$ regime the phase is approximately \cite{Futterman:1988ni,Glampedakis:2001cx}
\begin{align}
e^{2 i \delta_\ell} &\approx \frac{\Gamma(\ell + 1 - 2 i M \omega)}{\Gamma(\ell + 1 + 2 i M \omega)} \nn \\ 
&\approx \exp(-4 i M \omega \ln L) \times \left(1 + O(L^{-2}) \right)  \label{eq:asym-phase} , 
\end{align}
where $L = \ell + 1/2$.  In the large-$\ell$ regime, we may use a uniform asymptotic approximation for the Legendre polynomial,
\begin{equation}
P_\ell( \cos \theta ) \approx \sqrt{\frac{2}{\pi L \sin \theta}} \sin \left( \frac{\pi}{4} + L \theta \right) . \label{eq:asym-legP}
\end{equation}
Inserting Eqs.~(\ref{eq:asym-phase}) and (\ref{eq:asym-legP}) into the series (\ref{eq:f0schw}) yields
\begin{equation}
f_0(\theta) \approx \frac{1}{i \omega} \sqrt{\frac{2}{\pi \sin \theta}} \sum_{\ell} L^{1/2} \left( e^{-4 i M \omega \ln L} - 1 \right) \sin \left( \frac{\pi}{4} + L \theta \right) .
\end{equation}
The coefficients of the series on the right-hand side do not approach zero in the limit $L \rightarrow 0$, and thus this series is not convergent, i.e., it is divergent. Heuristically, one can see that this happens because the scattering coefficient $\mathcal{S}_{\lmws}^\pm - 1$, defined in Eq.~(\ref{eq:scat-coef}), remains of order unity even in the large-$l$ limit.

Similarly, in the Kerr context, the series representations of $f_s$ are also found to be divergent and thus impractical for use without modification.

\subsection{The physical origin of the divergence\label{subsec:physical}}
Heuristically, the poor convergence of the partial wave series (\ref{eq:amp_fs}) is due to the fact that the scattering amplitude $f_s(\theta, \phi)$ diverges at the antipode of the point on the celestial sphere which corresponds to the `centre' of the incident wave (see Fig.~\ref{fig:geometry}). The divergence in $|f_s(\theta,\phi)|$ scales as $\Theta^{-2}$ close to this antipodal point, where $\Theta$ is the angle on the sphere between the scattering direction $(\theta,\phi)$ and the antipode at $(\gamma,\phi_0)$. Thus, $d\sigma / d\Omega$ diverges as $\Theta^{-4}$, in the same manner as the Rutherford cross section in quantum-mechanical scattering. Ultimately, this divergence is due to the fact that gravity, like electromagnetism, is a long-ranged force with a potential that falls off as $1/r$ in the Newtonian limit. 

In the geometric-optics limit ($M\omega \gg 1$) the divergence in $d\sigma / d\Omega$ can be understood as follows: (i) rays passing through an annulus of radius $b \gg M$ and width $db$ on the incident wavefront (with area $2 \pi b db$) are deflected through the Einstein scattering angle $\Theta = 4M / b$; (ii) these rays are scattered into a solid angle $d \Omega = 2 \pi \sin \Theta d\Theta$; (iii) the classical scattering cross section, defined as the area on the wavefront divided by solid angle on the sphere, is then
\begin{align}
\left. \frac{d \sigma}{d\Omega} \right|_{cl} =  \frac{b}{\sin \Theta \left| \frac{d\Theta}{d b} \right|} , 
\end{align}
and (iv) inserting $\Theta = 4M / b \ll 1$, leads to
\begin{align}
\left. \frac{d \sigma}{d\Omega} \right|_{cl} \approx \frac{16 M^2}{\Theta^{4}} .
\end{align}

The issue facing a practical calculation is that the partial wave sum representation of $f_s$ (Eq.~(\ref{eq:amp_fs})) is not  convergent for \emph{any} value of $\theta$. This is not unexpected, as such behaviour is typical in Fourier series expansions of singular functions. Here we shall overcome this practical limitation by adapting a method that originates in a 1950s work on electron scattering \cite{Yennie:1954zz}.

\subsection{Series reduction\label{subsec:series-reduction}}
As discussed above, a physical divergence in the amplitude $f_s$ is expected at the antipodal point on the sphere at $(\gamma, \phi_0)$ in spherical polars. 
Taking $\phi_0 = \pi/2$ by convention \cite{Glampedakis:2001cx}, the angle $\Theta=\Theta(\theta,\phi)$ on the minor arc connecting the point $(\theta, \phi)$ to the antipodal point is defined by 
\begin{equation}
\cos \Theta  \equiv \cos \gamma \cos \theta +  \sin \gamma \sin \phi \sin \theta. \label{def:cosxi}
\end{equation}
Our aim is to `reduce' the divergence in the series at $\cos \Theta = 1$, by defining the $k^{\text{th}}$ \emph{reduced series} as
\begin{eqnarray}
f_s^{(k)}  &\equiv&   \left( 1 - \cos \Theta \right)^k f_s .
\label{eq:red_series_fs}
\end{eqnarray}
We show below that (i) the series coefficients for $f_s^{(k)}$ are found from certain linear combinations of the coefficients for $f_s$, and (ii) the series for $f_s^{(k)}$ has improved convergence properties, allowing a practical numerical calculation of $f_s^{(k)}$. The amplitude $f_s$ is then calculated from
\begin{equation}
f_s(\theta,\phi)  = \frac{f_s^{(k)}}{  \left( 1 - \cos  \Theta  \right)^k }.
\end{equation}


%

\subsection{From spheroidal to spherical harmonics\label{subsec:spheroidal}}

The spin-weighted spheroidal harmonics $\sh(\theta)$ featuring in Eqs.~(\ref{eq:amp_fs}) and (\ref{eq:amp_g2}) satisfy the angular Teukolsky equation \cite{Teukolsky:1972my}, namely
\begin{align}
\frac{1}{\sin \theta} \dth \left( \sin \theta \frac{ d \sh}{d \theta} \right) + \bigg( & a^2 \omega^2 \cos^2 \theta - \frac{m^2}{\sin^2 \theta}  \nonumber \\ 
& \quad - \frac{2ms \cos \theta }{\sin^2 \theta } - 2 a \omega s \cos \theta - s^2 \cot^2 \theta + s + A_{lm} \bigg) \sh = 0 .
\end{align} 
In the limit $a\omega = 0$, the functions $\sh (\theta) e^{im \phi}$ reduce to spin-weighted \emph{spherical harmonics}, $\sylm(\theta) e^{im\phi}$ \cite{Goldberg:1966uu}. 

Any spin-weighted \emph{spheroidal} harmonic (indeed any well-behaved function on the sphere \cite{Newman:1966ub}) may be expanded in the basis of spherical harmonics of the same spin weight \cite{Hughes:1999bq, Glampedakis:2001cx, Dolan:2008kf}, viz.,
\begin{equation}
\sh (\theta) = \sum_{j= \text{max} \{ |m|, |s| \} }^{\infty} b_{jm \ell}^s \, \syjm (\theta) ,
\end{equation}
where $b_{jm \ell}^s$ are series coefficients. In practice only a few coefficients $b_{jm\ell}^s$ are typically required, as they exhibit an exponential fall-off for $|j - \ell| \gg 1$ \cite{Glampedakis:2001cx}.

We now define
\begin{equation} \label{def:fml}
f_{\ell m}^s \equiv \frac{2 \pi}{i \omega} \sh (\gamma) e^{-i \phi_0} \widehat{\mathcal{S}}_{\lmws} , \quad   |m| \leq \ell,
\end{equation}
where $ \widehat{\mathcal{S}}_{\lmws} = \frac{1}{2} \left( \mathcal{S}^{+}_{\lmws} + \mathcal{S}^{-}_{\lmws} \right)$
and 
\begin{equation} \label{def:Fjm}
F_{jm}^s \equiv \sum_{\ell = |m|}^{\infty} f_{\ell m}^s b_{jm \ell}^s ,
\end{equation}
so that the scattering amplitude (\ref{eq:amp_fs}) can be written in the form
\begin{equation} \label{eq:ampp_fs_sphericalharm_exp}
f_s \angarg = \sum_{j=s}^{\infty} \sum_{m=-j}^{j} F_{jm}^s \, \syjm (\theta) e^{i m \phi}.
\end{equation}
%

\subsection{Regulating the series\label{subsec:regulation}}
Here, and in the following sections, it will be necessary to move a factor inside infinite series in $j$. Strictly, such a step is invalid for divergent infinite series. However, we may evade this issue by taking as our starting point a regulated sum $f^{(\epsilon)}_s$ that is convergent for $\epsilon > 0$, i.e.,
\begin{equation}
f^{(\epsilon)}_s \angarg =  \sum_{j=s}^{\infty} \sum_{m=-j}^{j} F_{jm}^s  \syjm (\theta) e^{i m \phi} \, \Xi_{\epsilon}(j), \label{eq:regulated-series}
\end{equation}
where here $\Xi_\epsilon(\cdot)$ is a regulating function introduced to smoothly cut off the infinite sum, such that (for $\epsilon > 0$) $\Xi_\epsilon(x) \rightarrow 0$ sufficiently rapidly that the series (\ref{eq:regulated-series}) is convergent. The family of functions $\Xi_\epsilon(x)$ should be such that
\begin{equation}
f_s \angarg = \lim_{\epsilon \rightarrow 0} \, f_s^{(\epsilon)} (\theta, \phi) ,
\end{equation}
and we take this limit at the end of the process.  An example of a regulating factor is $\Xi_\epsilon(x) = \frac{1}{2} \left( \tanh( 1/\epsilon - x ) + 1 \right)$. 

The physical motivation underpinning the above is that, in practice, a divergent series (\ref{eq:amp_fs}) is a consequence of starting with a planar wave of infinite extent; and by introducing a smooth cut-off to the sum we can limit the extent of the initial wavefront in a controlled manner. For clarity, we shall not include the regulating factor in any of the steps below, and we implicitly take the limit $\epsilon \rightarrow 0$ at the end of the process.

\subsection{Scalar field case\label{subsec:s0}}
We start with the scalar field case $s = 0$. 
Let us define $G_{jm} \equiv A_{jm} F_{jm}^0 $, where $A_{jm}$ is given in Eq.~(\ref{eq:Alm}), so that 
\begin{equation}\label{eq:scat_scalar_Kerr}
f_0 (x,\phi) =   \sumjms G_{jm}  P_{j}^m(x) e^{im\phi} ,
\end{equation}
where 
\begin{equation}
x = \cos \theta,
\end{equation}
and
$\sumjms$ is a shorthand for $\sumjm$. Here we have used Eq.~(\ref{def:ylm}) to rewrite the (scalar) spherical harmonics in terms of associated Legendre polynomials $P_{j}^m(\cdot)$. 
Now we define the $k_{\text{th}}$ reduced series $f^{(k)}_0$ and its coefficients $G_{jm}^{(k)}$ in accordance with Eq.~(\ref{eq:red_series_fs}), that is,
\begin{align}
f^{(k)}_0 (x,\phi) &\equiv \big( 1 -\cos (\Theta ) \big)^k f_0(x,\phi) \label{def:kth_red_ser}  \\
 & = \sumjms G_{jm}^{(k)}  P_{j}^m(x) e^{im\phi} , \label{def:kth_red_coeff}
\end{align}
recalling that $\Theta(\theta, \phi)$, defined in Eq.~(\ref{def:cosxi}), is the angle to the antipodal point.
It is useful at this point to express $\cos \Theta$ in terms of $x \equiv \cos \theta$ and $\phi$ as 
\begin{equation}
\cos \Theta = x \cos \gamma +  \frac{1}{2i} \sqrt{1 - x^2} \sin(\gamma) (e^{i\phi} - e^{-i\phi}). \label{def:cosxi3}
\end{equation}
To find the recursion relation for $G_{jm}^{(k)}$ we make the argument
\begin{subequations}
\begin{align}
f^{(k+1)}_0 (x,\phi) &=  \l( 1 -\cos \Theta \r) \sumjms G_{jm}^{(k)}  P_{j}^m(x) e^{im\phi}  \\
&= \sumjms \l( 1 -\cos \Theta \r) G_{jm}^{(k)}  P_{j}^m(x) e^{im\phi} \\
& = \sumjms \l( 1 - x \cos \gamma -  \frac{1}{2i} \sqrt{1 - x^2} \sin(\gamma) (e^{i\phi} - e^{-i\phi}) \r) G_{jm}^{(k)}  P_{j}^m(x) e^{im\phi}. \label{eq:eq:rec_step1}
\end{align}
\end{subequations}
We now set $G_{jm}^{(k)} =0$, for $j<|m|$ and $j<0$, in order to write the sums below in a compact fashion. Using the recursion relations for associated Legendre polynomials given in Eqs.~(\ref{eqs:leg_recs}) and~(\ref{eqs:leg_recs_init}) of the Appendix, we establish that
%
%
%
%
\begin{subequations}
\begin{align}
\sumjms \sqrt{1-x^2} e^ {+i \phi} G_{jm}^{(k)}  P_{j}^m(x) e^{im\phi} &=  \sumjms \frac{G^{(k)}_{(j+1)(m-1)}}{2j+3} P_{j}^{m}(x) e^{i m\phi} -  \frac{G^{(k)}_{(j-1)(m-1)} }{2j-1}  P_{j}^{m}(x)   e^{i m\phi}  , \label{eq:epphi_red} \\
\sumjms \sqrt{1-x^2} e^ {-i \phi} G_{jm}^{(k)}  P_{j}^m(x) e^{im\phi} & = \sumjms G^{(k)}_{(j-1)(m+1)} \frac{(j-m-1)(j-m)}{2j-1}  P_{j}^{m}(x) e^{im\phi} \nonumber \\
& \phantom{000} -  G^{(k)}_{(j+1)(m+1)} \frac{(j+m+1)(j+m+2)}{2j+3}  P_{j}^{m}(x)  e^{i m\phi} , \label{eq:emphi_red} \\ 
\sumjms x G_{jm}^{(k)}  P_{j}^m(x) e^{im\phi} &= \sumjms \left[ \frac{j-m}{2j-1} G^{(k)}_{(j-1)m} + \frac{j+m+1}{2j+3} G_{(j+1)m}^{(k)} \right] P_{j}^{m}  e^{im\phi}. \label{eq:cosfk_red_factor}
\end{align}
\end{subequations}
Substituting Eqs.~(\ref{eq:epphi_red}) to (\ref{eq:cosfk_red_factor}) 
into Eq.~(\ref{eq:eq:rec_step1}), gives the recursion relation  \\
\begin{align}
G^{(k+1)}_{jm} &= G^{(k)}_{jm} - \cos \gamma \left[  \frac{j-m}{2j-1} G^{(k)}_{(j-1)m} + \frac{j+m+1}{2j+3} G_{(j+1)m}^{(k)} \right]  \nonumber\\[5pt]
& \phantom{=} - \frac{1}{2i} \sin \gamma \bigg[ - \frac{(j-m-1)(j-m)}{2j-1} G^{(k)}_{(j-1)(m+1)} - \frac{1}{2j-1}  G^{(k)}_{(j-1)(m-1)}  \nonumber \\[5pt]
& \hspace{2.0cm}  + \frac{1}{2j+3} G^{(k)}_{(j+1)(m-1)} +  \frac{(j+m+1)(j+m+2)}{2j+3} G^{(k)}_{(j+1)(m+1)}   \bigg] \label{eq:recursion_Glm}.
\end{align}
%
The scattering amplitude $f_0$ is then computed using
\begin{equation}
f_0 \angarg = \frac{1}{(1 - \cos \Theta)^k} \sumjms \frac{G_{jm}^{(k)}}{A_{jm}} Y_{jm}(\theta) e^{i m \phi} . \label{eq:scat_amp_s0}
\end{equation}

\subsection{Electromagnetic case\label{subsec:s1}}
We now proceed to the electromagnetic case ($s = 1$). 
Define $G_{jm}^1 \equiv - A_{jm} F_{jm}^1 / (\sqrt{j(j+1)} $, where $F_{jm}^1$ and $A_{jm}$ are defined in Eq.~(\ref{eq:Alm}) and Eq.~(\ref{def:Fjm}), respectively, so that 
\begin{equation}
f_1 (x,\phi) =   \check{\delta}_{0} \ff_1 (x,\phi),
\end{equation}
where
\begin{equation}
\ff_1 (x,\phi) = \sumjms G_{jm}^1  P_{jm}(x)  e^{im\phi},
\end{equation}
and $\check{\delta}_{0}$ is the spin lowering operator defined in Eq.~(\ref{def:ylms}).
Next, we define the $k_{\text{th}}$ reduced series $\ff^{(k)}_1$ and its coefficients $G_{jm}^{1(k)}$, in accordance with Eq.~(\ref{eq:red_series_fs}), that is,
\begin{align}
\ff^{(k)}_1 (x,\phi) &\equiv \big( 1 -\cos \Theta \big)^k \ff_1(x,\phi) \label{def:kth_red_ser_s1} \\
& =   \sumjms  G_{jm}^{1(k)}  P_{jm}(x)  e^{im\phi}. \label{def:kth_red_coeff_s1}
\end{align}
Here we have moved the spin operator outside the summation in the second line. 
Proceeding recursively, it is now clear that for $k\geq1$, the $G_{jm}^{1(k)}$ can be calculated with exactly the same recursion relation as in the $s=0$ case, that is, Eq.~(\ref{eq:recursion_Glm}) with $G_{jm}^{(k)}$ replaced by $G_{jm}^{1(k)}$. 
The amplitude $f_1$ is then calculated with the expression
\begin{align}
f_1 (\theta,\phi) & = \sumjms \Bigg[  \l( \check{\delta}_{0} \l[ \frac{1}{(1 - \cos \Theta)^{k}} \r] \r) Y_{jm}(\theta) - \frac{\sqrt{j(j+1)}}{(1 - \cos \Theta)^{k}}   \, {}_{-1} Y_{jm}(\theta) \Bigg] \frac{G_{jm}^{1(k)}}{A_{jm}} e^{im\phi} . 
\end{align}

%
%
%
%
\subsection{Gravitational wave case\label{subsec:s2}}
Define $G_{jm}^{2} \equiv - A_{jm} F_{jm}^2 / (\sqrt{(j-1)j(j+1)(j+2)})$, so that 
\begin{equation}
f_2 (x,\phi) =    \check{\delta}_{-1} \check{\delta}_{0} \ff_{2} (x,\phi),
\end{equation}
where
\begin{equation}
\ff_{2} (x,\phi) \equiv  \sumjms G_{jm}^{2}  P_{jm}(x)  e^{im\phi}
\end{equation}
and $\check{\delta}_{s}$ are the spin lowering operators (see Eq.~(\ref{def:ylms})).
Define the $k_{th}$ reduced series, and coefficients $G_{jm}^{2(k)}$, by
\begin{align}
\ff^{(k)}_2 (\theta,\phi) &\equiv \big( 1 -\cos \Theta \big)^k \ff_2(\theta,\phi) \label{def:kth_red_ser_s2} \\
& =   \sumjms  G_{jm}^{2(k)}  P_{jm}(x)  e^{im\phi}. \label{def:kth_red_coeff_s2}
\end{align}
Proceeding recursively, it is now clear that for $k\geq1$, the $G_{jm}^{2(k)}$ can be calculated with exactly the same recursion relation as in the $s=0$ case, that is, Eq.~(\ref{eq:recursion_Glm}) with $G_{jm}^{(k)}$ replaced by $G_{jm}^{2(k)}$. 
The amplitude $f_2$ is then calculated with the expression
\begin{multline} 
f_2 (\theta,\phi) = \sumjms \Bigg[  \l( \check{\delta}_{-1} \check{\delta}_{0} \l[ \frac{1}{(1 - \cos \Theta)^{k}} \r] \r) Y_{jm}(\theta) \\ - \frac{\sqrt{(j-1)j(j+1)(j+2)}}{(1 - \cos \Theta)^{k}}   \, {}_{-2} Y_{jm}(\theta) \Bigg] \frac{G_{jm}^{2(k)}}{A_{jm}} e^{im\phi} . 
\end{multline}
The process outlined above could be generalised to all integer values of $s$ by using further spin lowering (or raising) operators. 

%
%
%

\section{Results and discussion\label{sec:results}}
Here we present some results for scalar off-axis scattering to verify our method. 

\subsection{Estimating the error of truncation}\label{subsec:error}

In this section we address the question of how we can be confident that the above method does indeed result in a convergent sum. It is possible to show analytically, that for the comparison Newtonian problem on a Schwarzschild BH ($a\omega=0$), the reduced summations for $k \geq 1$ are convergent at all angles, except the antipodal point (see App.~(\ref{ap:convergence_schw})). The general problem is of course more difficult, since we have a summation over $m$ as well as $l$. However, in the large $l$ regime we could expect the spin of the BH to have a small effect on the phase shifts and thus would expect similar convergence properties for the reduced series. Here we will investigate the effects of series reduction for a few examples to test this assertion.

\begin{figure}
\centering
  \includegraphics[width=0.8\textwidth]{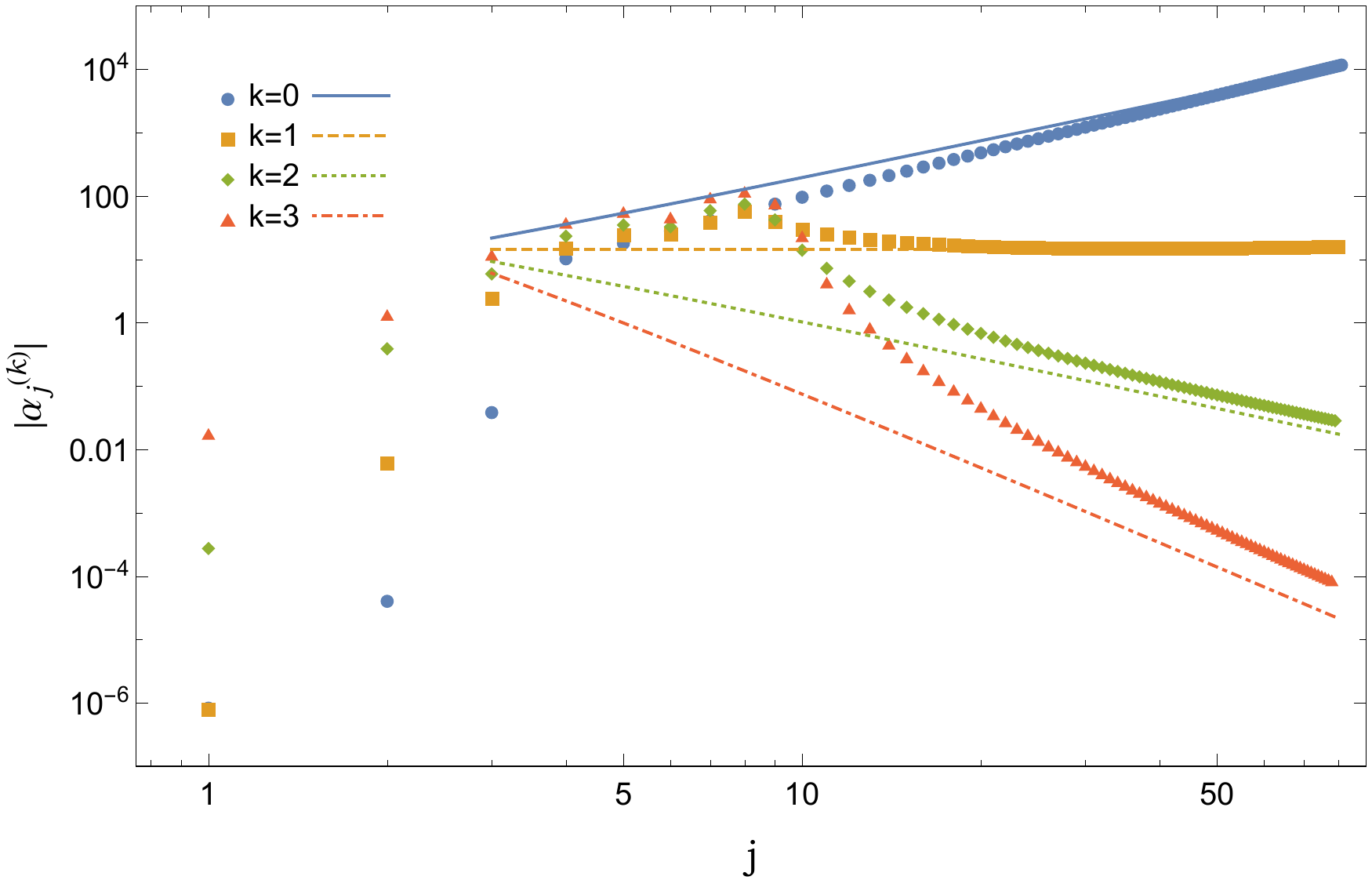}
\caption{ Absolute value of the $k^{\text{th}}$-reduced series terms $\alpha_{j}^{s(k)}$ against (integer) $j$. Here we have chosen the parameters $s=0$, $a = 0.9 M$, $\omega M=1$, $\theta_0=\gamma=\pi/2$, and $\phi_0=\pi/2$. For comparison we also plot lines $\propto (j+1/2)^{2-2k} $ for $k \in \{0,1,2,3 \}$.}
\label{fig:error_plt1}
\end{figure}

The addition theorem for spin-weight spherical harmonics implies
\begin{equation}
\sum_{m=-j}^{j} \left| \syjm \angarg \right|^2 = (-1)^s \sqrt{\frac{2j+1}{4\pi}} Y_{js}^{-s} (0,0) .
\end{equation}
(This is a special case of the theorem given by \cite{Michel2019}). From Eq.~(3.1) in \cite{Goldberg:1966uu}, and since $Y_{jm}^{-s} \angarg = Y_{jm}^{s} (\pi-\theta, \pi-\phi)$, it follows that $Y_{js}^{-s} (0,0) = (-1)^s \sqrt{(2j+1)/4\pi}$, and hence
\begin{equation} \label{eq:syjm_sum_m}
\sum_{m=-j}^{j} | \syjm \angarg |^2 =  \frac{(2j+1)}{4\pi} .
\end{equation}
It is interesting that the RHS of Eq.~(\ref{eq:syjm_sum_m}) is independent of the spin. Now, define
\begin{align}
F_{jm}^{s(k)} \equiv G_{jm}^{s(k)} / A_{jm}, \\
F_{j}^{s(k)} = \text{Max}\{ |F_{jm}^{s(k)} \}_{|m|<j }.
\end{align}
Then, it follows from the triangle and Cauchy-Schwarz inequalities that
\begin{equation} \label{eq:red_amp_ineq}
\left| \sum_{j=n}^{N} \sum_{m=-j}^{j} F_{jm}^{s(k)}  Y_{jm} \angarg \right| \leq \sum_{j=n}^{N}  \alpha_{j}^{s(k)} ,
\end{equation} 
where
\begin{equation} \label{def:leadingerrorterm}
\alpha_{j}^{s(k)} = \left| \frac{(2j+1)^2}{\sqrt{4\pi}} F_{j}^{s(k)} \right|.
\end{equation}
If we truncate the summation for calculating $\ff^{(k)}_s$  in Eq.~(\ref{def:kth_red_coeff}), ~(\ref{def:kth_red_coeff_s1}) or (\ref{def:kth_red_coeff_s2}) at some $j=J$, then the absolute error is bounded by the RHS of Eq.~(\ref{eq:red_amp_ineq}). If the sequence ($\alpha_{j}^{s(k)}$) is decreasing for $j>J$, then $\alpha_{J+1}^{s(k)}$ gives us a reasonable estimate of the error. We present plots of $\alpha_{j}^{0(k)} $ against $j$ for $\omega M=1$ and $a =0.9M$ in Fig.~\ref{fig:error_plt1}. The numerical evidence suggests that 
%
\begin{equation}
\alpha_{j}^{0(k+1)} = O (  \alpha_{j}^{0(k)}/j^2 )  \quad , \quad \text{as } l \rightarrow \infty. 
\end{equation}
This might be expected as the original reduction process, which we have based our method on, showed the equivalent property (Eq.~(50) in \cite{Yennie:1954zz}). In App.~(\ref{ap:convergence_schw}) we show how this improvement in the summation convergence can be proven explicitly for the special case of no rotation and $s=0$ (and assumption that the phase shift tends to the comparison Coulomb value).

In Fig.~\ref{fig:error_plt1} we see that the error bound $\alpha_{j}^{0(3)}$ is negligible for $j \gtrsim 50$ (when $\omega M=1$ and $a=0.9M$). More terms in the series are needed to reduce the error to a desired level if we increase $\omega M$ or $a/M$.   
The numerical evidence and proof of convergence for the comparison Newtonian problem (App.~(\ref{ap:convergence_schw})) are, we think, sufficient evidence to be confident in our final scattering cross section calculations.


\subsection{Differential scattering cross sections: scalar case}\label{subsec:csec}

Here we present a selection of our results for the scalar case, computed using the series reduction method.
The numerical method we use for calculating phase shifts can be found in Ref.~\cite{Futterman:1988ni} (see also Ref.~\cite{Glampedakis:2001cx} for a method based on the Pr{\"u}fer transformation).

\begin{figure*}
\centering
  \includegraphics[width=0.5\textwidth]{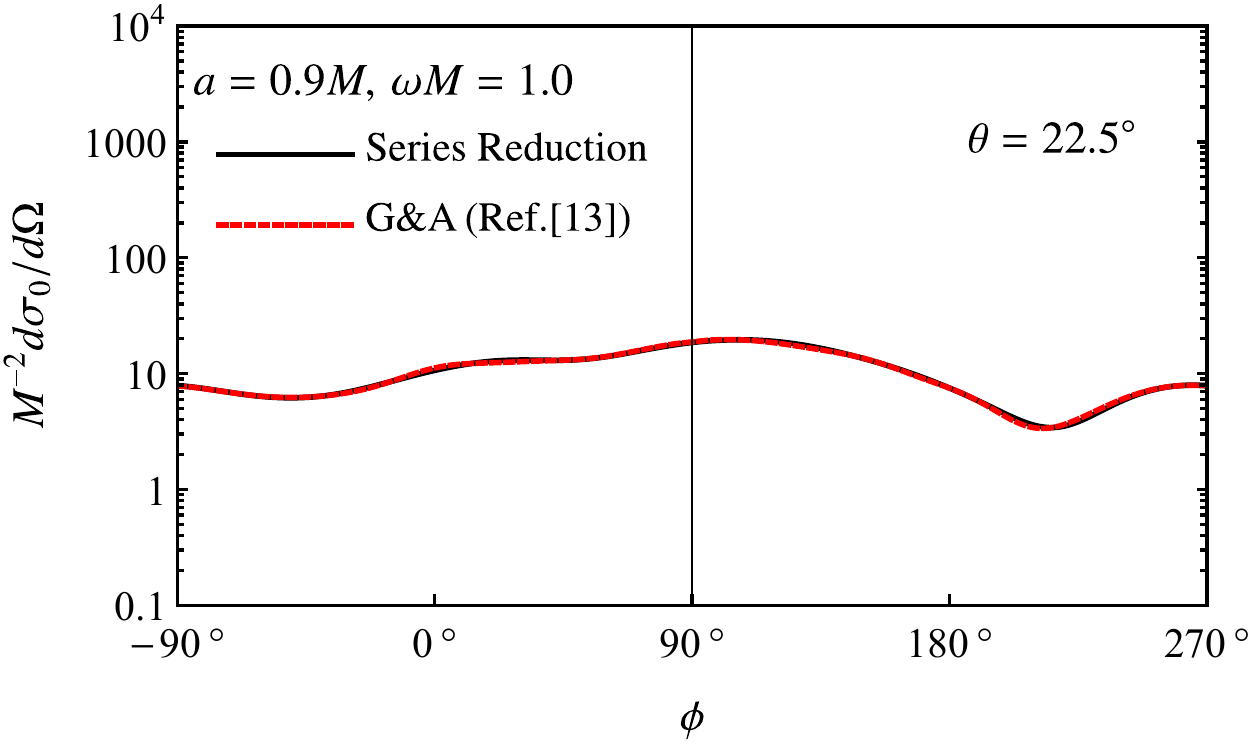}\includegraphics[width=0.5\textwidth]{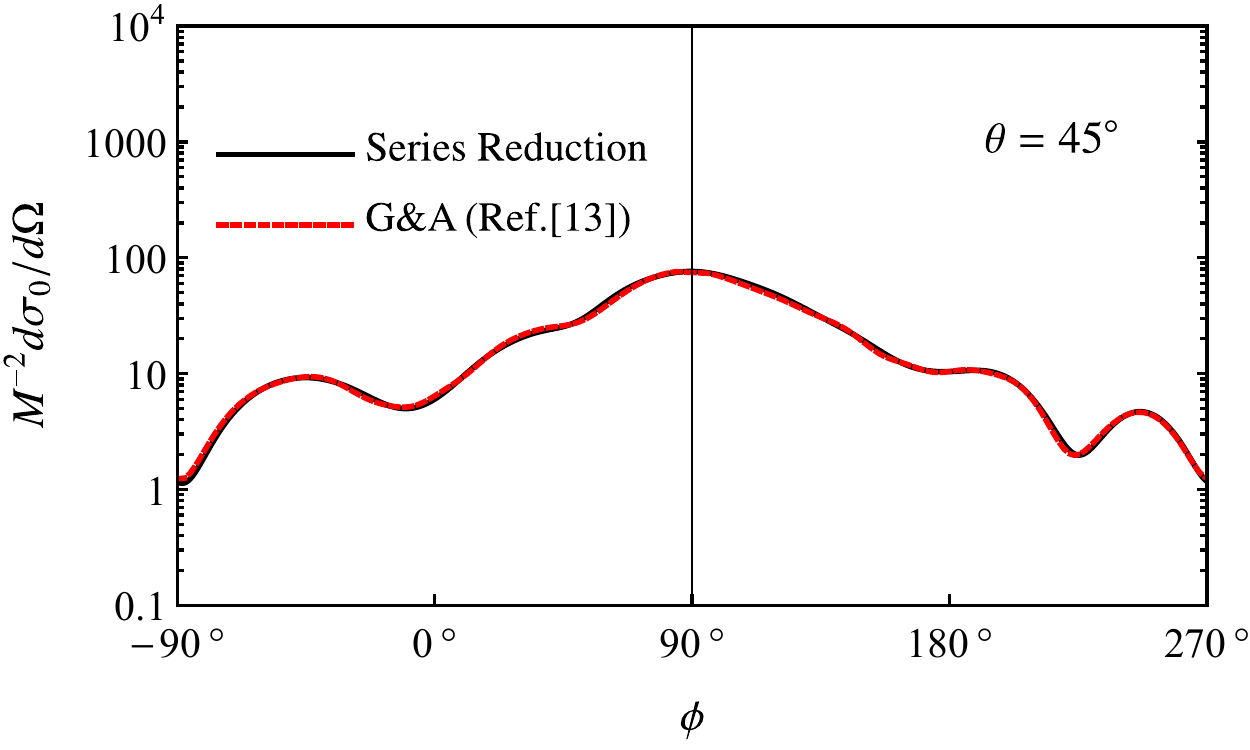}\\
    \includegraphics[width=0.5\textwidth]{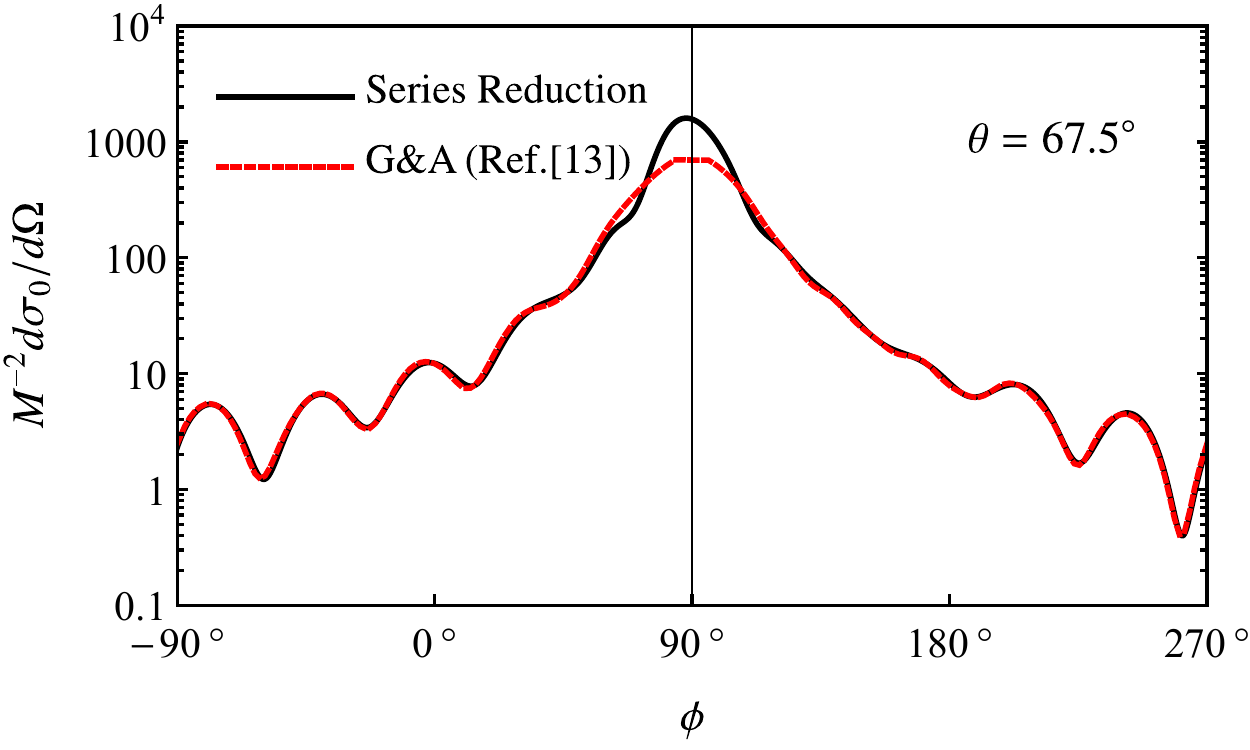}\includegraphics[width=0.5\textwidth]{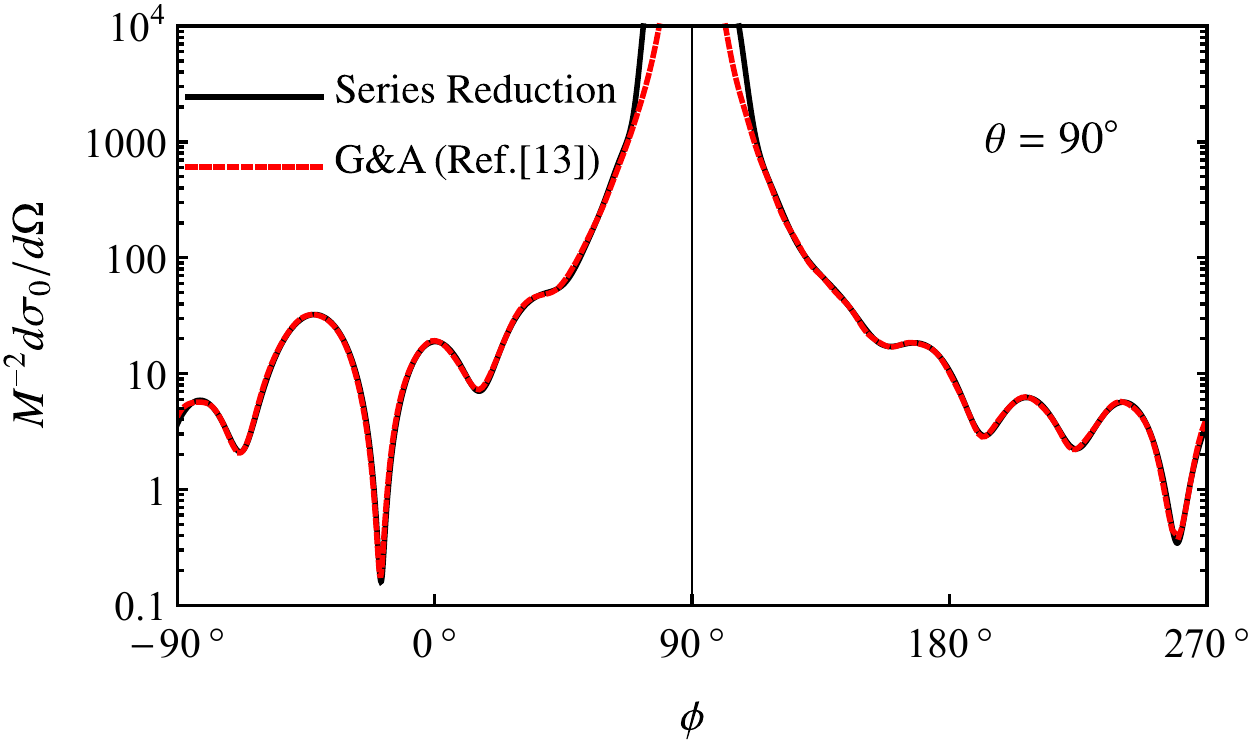}
\caption{Off-axis differential scattering cross section for scalar plane waves with $\omega M=1.0$, impinging upon a Kerr BH with $a=0.9M$. The vertical line represents the forward direction~($\gamma=\phi_0=90^{\circ}$). The results of Glampedakis and Andersson (G{\&}A) from Ref.~\cite{Glampedakis:2001cx} are shown in dashed red for comparison.
} 
\label{fig:cross_sections}
\end{figure*}

Figure~\ref{fig:cross_sections} exhibits the differential scattering cross sections as functions of $\phi$~($-90^{\circ}<\phi<270^{\circ}$), for fixed values of $\theta$~($\theta=22.5^{\circ}$, $45^{\circ}$, $67.5^{\circ}$, and $90^{\circ}$). The incidence direction of the scalar waves~(with $\omega M=1.0$) lies on the equatorial plane~($\gamma=90^{\circ}$, $\phi_0=90^{\circ}$) of the rotating Kerr BH~($a=0.9M$). 

We compare our results, computed via the series reduction method~(with the scattering amplitude given by Eq.~\eqref{eq:scat_amp_s0}), with those presented in the top panel of Fig. $9$ of Ref.~\cite{Glampedakis:2001cx}, computed by splitting the scattering amplitude into `Newtonian' and diffraction amplitudes. A good agreement can be observed among the results obtained via the series reduction with those shown in Ref.~\cite{Glampedakis:2001cx}.
For the plots exhibited in Fig.~\ref{fig:cross_sections}, we have terminated the summation (Eq.~\eqref{eq:scat_amp_s0}) at $l_{max}=30$, $j_{max}=18$, and used $k=3$ applications of the reduction algorithm. 

The main features in the cross sections shown in Fig.~\ref{fig:cross_sections} are: a forward Coulomb divergence (at $\theta=\phi=90^{\circ}$, see bottom right panel); an asymmetry with respect to the direction of incidence (indicated by the vertical lines);  a non trivial dependence on the polar and azimuthal angle of observation; and a glory maximum for the equatorial plane cross section ($\theta=90^{\circ}, \phi \approx -42^{\circ}$, bottom right panel). For a more detailed discussion of the cross section features and their interpretation, we refer the reader to Ref.~\cite{Glampedakis:2001cx}.

\section{Discussion}\label{sec:conclusions}

In this work we have devised a method to overcame a significant obstacle in computing scattering cross sections for bosonic plane waves impinging on a Kerr BH with an arbitrary angle of incidence. Namely, the divergence of the partial wave scattering amplitude sum in its standard formulation has been ameliorated by extending the series reduction method devised by Yennie~\emph{et al.} (and extended to BH scattering in Refs.~\cite{Dolan:2006vj,Dolan:2008kf}). 

We have demonstrated the validity of the series reduction method for scalar plane waves, when applied to scattering scenarious with $\omega M \sim 1$ (where diffraction effects are most prominent) and where the BH may be rapidly spinning ($a \approx M$). We have compared our results with those obtained by Glampedakis and Andersson using an alternative `Newtonian splitting' method \cite{Glampedakis:2001cx}. The results show good agreement except for scattering angles $(\theta,\,\phi)$ near to the forward direction $(\gamma,\,\phi_0)$, where the cross section diverges and numerical errors become hard to control. This can be overcome by including more terms in the partial wave expansions. For all other angles in the plots exhibited, our method shows good convergence of the scattering-amplitude reduced series (see Fig.~\ref{fig:error_plt1}), and agrees well with the Newtonian splitting method. 



In addition, we have given a proof of convergence for the reduced series for a scalar wave incident on a Schwarzschild BH (App.~(\ref{ap:convergence_schw})). Whether this proof generalises to a Kerr BH and arbitrary angle of incidence is an open question. Intuitively we expect it to, since the rotation of the BH has a negligible effect on partial waves of a sufficiently large mode number. In practice proving this would be difficult.  
However, given the preliminary results we believe the method can provide accurate results for general off axis scattering of bosonic fields. 


In related work, Folacci and Ould El Hadj have shown that Schwarzschild BH scattering cross sections can be accurately calculated using complex angular momentum techniques, as opposed to partial wave series expansions \cite{Folacci:2019vtt,Folacci:2019cmc}. They plan to consider Kerr BHs in the future. This will be a particularly interesting challenge since the introduction of rotation both promotes the role of angular momentum in any physical processes, and obscures the path to developing useful complex analysis tools to understand them.

Here we have given little in the way of physical interpretation (see however \cite{Glampedakis:2001cx}), instead focusing on the computational method. In a more detailed work to follow we aim to remedy this. For example, it is known that perturbations incident on a Kerr BH may exhibit superradiance - an amplification due to extraction of the BHs rotational energy. This is spin dependent, and can be particularly strong in the GW case: Teukolsky and Press found a maximum superradiant amplification of $138 \% $ for the $l=m=2$ mode when $a=0.99999M$ and $\omega = 2 \omega^+ $ (where $\omega^+ \equiv a/(2Mr_+)$ is the `angular velocity of the horizon') \cite{Teukolsky:1974yv}. The implications of superradiance for monochromatic off-axis scattering are yet to be fully explored \cite{Handler:1980un,Glampedakis:2001cx}. This is of foundational interest to provide a full understanding of the BH superradiance phenomenon, which may have observational consequences \cite{Arvanitaki:2010sy}.


\section*{Acknowledgements}

We thank Kostas Glampedakis and Nils Andersson for sharing their data to facilitate a comparison of results. TS acknowledges financial support from the Engineering and Physical Sciences Research Council (EPSRC) and the University of Sheffield Publication Scholarship. LCBC and LCSL would like to acknowledge Conselho Nacional de Desenvolvimento Cient\'ifico e Tecnol\'ogico (CNPq) and Coordena\c{c}\~ao de Aperfei\c{c}oamento de Pessoal de N\'ivel Superior (CAPES) -- Finance Code 001, from Brazil, for partial financial support.
SD~acknowledges additional financial support from the Science and Technology Facilities Council (STFC) under Grant No.~ST/P000800/1.
This research has also received funding from the European Union's Horizon 2020 research and innovation programme under the H2020-MSCA-RISE-2017 Grant No. FunFiCO-777740.

\appendix

\section{Spherical harmonics and the spin-lowering operator}

Spherical harmonics can be defined in terms of associated Legendre polynomials

\begin{equation} \label{def:ylm}
Y_{lm} \angarg \equiv A_{lm} P_l^{m}( \cos \theta) e^{i m \phi}, 
\end{equation}
where
\begin{equation}
A_{lm} \equiv \sqrt{\frac{2l+1}{4\pi} \frac{(l-m)!}{(l+m)!} }. \label{eq:Alm}
\end{equation}
The operator $\check{\delta}_s$ lowers the spin of a harmonic \cite{Goldberg:1966uu,Newman:1966ub}
\begin{align} \label{def:ylms}
\check{\delta}_s Y_{lm}^{s} &= - (\sin \theta)^{-s} \left[ \frac{ \p }{ \p \theta} - \frac{i}{\sin \theta} \frac{ \p }{ \p \phi} \right] (\sin \theta)^s Y_{lm}^{s} \\
& = - [(l+s)(l-s+1)]^{1/2} Y_{lm}^{s-1}, 
\end{align}
thus
\begin{subequations} \label{eq:neg_spin_harms}
\begin{align}
Y_{lm}^{-1} &= -[(l)(l+1)]^{-1/2} \check{\delta}_0 Y_{lm}, \\ 
Y_{lm}^{-2} &= [(l-1)l(l+1)(l+2)]^{-1/2} \check{\delta}_1 \check{\delta}_0  Y_{lm}. 
\end{align}
\end{subequations}
In turn, the associated Legendre polynomials are 
\begin{equation}
P_l^{m}( \cos \theta) \equiv (-1)^m (1-\cos^2 \theta)^{m/2} \frac{d^m}{d (\cos \theta) ^m} \left[ P_l( \cos \theta) \right] ,
\end{equation}
and they satisfy
\begin{equation}\label{eq:asleg_def}
(1-x^2) \frac{d^2 P_l^m}{dx^2} - 2x  \frac{dP_l^m}{dx} + \left( l(l+1) - \frac{m^2}{1-x^2} \right) P_l^m = 0.
\end{equation}

\section{Recursion relations for associated Legendre polynomials}

Some useful recursion relations for associate Legendre polynomials are
\begin{subequations} \label{eqs:leg_recs}
\begin{align}
 \sqrt{1-x^2}P_l^m &= \frac{1}{2l+1} \left( - P_{l+1}^{m+1} + P_{l-1}^{m+1} \right) \label{eq:leg_rec_m+}, \\
  \sqrt{1-x^2}P_l^m &= \frac{1}{2l+1} \left( (l-m+1)(l-m+2) P_{l+1}^{m-1} - (l-m+1)(l+m) P_{l-1}^{m-1} \right) \label{eq:leg_rec_m-}, \\
  x P_l^m & = \frac{1}{2l+1} \left( (l-m+1)P_{l+1}^{m} + (l+m) P_{l-1}^m \right)  \label{eq:leg_rec_l+-}.
\end{align}
\end{subequations}
Initial values for the first recursions of Eq.~(\ref{eqs:leg_recs}):
\begin{subequations}\label{eqs:leg_recs_init}
\begin{align}
 \sqrt{1-x^2}P_l^l &= -\frac{1}{2l+1}  P_{l+1}^{l+1}, \\
 \sqrt{1-x^2}P_l^{l-1} &= -\frac{1}{2l+1}  P_{l+1}^{l}. 
\end{align}
\end{subequations}
For the first derivative we make use of 
\begin{equation} \label{eq:leg_rec_deriv1}
(1-x^2) \frac{dP_l^m}{dx} = \frac{1}{2l+1} \left( (l+1)(l+m)P_{l-1}^m - l(l-m+1) P_{l+1}^m \right). 
\end{equation}
%

\section{Convergence of the Schwarzschild scattering amplitude series} \label{ap:convergence_schw}

Without loss of generality we can choose $\gamma=0, \phi_0 = \pi/2$, which implies 
\begin{equation}\label{eq:Glm_Schw}
G_{lm} = \delta_{m0} (2l+1)(e^{2i\delta_l}-1)/(2i\omega),
\end{equation}
and thus recovers Eq.~(\ref{eq:f0schw}) from Eq.~(\ref{eq:scat_scalar_Kerr}). In this case we only need to deal with a sum over $l$. It is convenient to switch variable from $l$ to $\lambda \equiv l+1/2$. Defining $b_\lambda^{(k)} \equiv G_{l0}^{(k)}$, we see from Eq.~(\ref{eq:recursion_Glm}) that 
\begin{equation}
b_{\lambda}^{(k+1)} = b_{\lambda}^{(k)} - \frac{1}{2} \left[ b_{\lambda+1}^{(k)} + b_{\lambda-1}^{(k)} \right] + \frac{1}{4} \left[ (\lambda+1)^{-1} b_{\lambda+1}^{(k)}   - (\lambda-1)^{-1} b_{\lambda-1}^{(k)}   \right].   
\end{equation}  
Suppose that in the large $l$ ($\lambda$) limit,
\begin{equation}
b_{\lambda}^{(k)} \sim \lambda^p \sum_{n=0}^{\infty} \alpha_n \lambda^{-n}, 
\end{equation}
then it follows from Eq.~(\ref{eq:recursion_Glm}) that 
\begin{align}\label{eq:asymptotic_m+1}
b_{\lambda}^{(k+1)} & \sim - \lambda^p \sum_{n=0}^{\infty} \alpha_n \lambda^{-n}     \sum_{j = 1}^{\infty} \left[  {p-n \choose 2j}  +  \frac{1}{2} { p-n-1 \choose 2j} \right] \lambda^{-2j}  \\
& \sim   \lambda^p \left[ \frac{1}{2} (p-1)^2 \alpha_0 \lambda^{-2} + O(\lambda^{-3}) \right].
\end{align}
If $p \neq 1$, then this implies 
\begin{equation}
\left| \frac{ G_{l0}^{(k+1)} }{ G_{l0}^{(k)} } \right| = l^{-2}, 
\end{equation}
so the series needs to be reduced at least $k> \text{Re}\{p\}/2$ times (assuming $\text{Re}\{p\}>0$), in order for it to converge for $\theta \neq 0$ (this can be seen by noting that $|P_l(\cos \theta)|<1$ for $\theta \neq 0, \pi$ and $P_l(-1) = (-1)^{l}$, then applying the ratio test and alternating series test for convergence, respectively). One can split the RHS of Eq.~(\ref{eq:Glm_Schw}) into two terms,
\begin{equation}\label{eq:Glm_Schw_2term}
G_{lm} = \delta_{m0} (2l+1)e^{2i\delta_l}/(2i\omega) - \delta_{m0} (2l+1)/(2i\omega).
\end{equation}
Setting $b_\lambda^{(k)} \equiv (2l+1)e^{2i\delta_l}/(2i\omega)$, and using Eq.~(\ref{eq:asym-phase}), we find $p=1-4iM\omega$. Choosing $b_\lambda^{(k)} \equiv (2l+1)/(2i\omega)$ gives $p=1$, and the series reduction method applied to this term will accelerate convergence even faster (in fact this sum is zero for $\theta \neq 0$). For convergence then, we must reduce the series at least twice (this is confirmed with numerical results). 

\bibliography{offaxisKerr}
\bibliographystyle{apsrev4-1}

\end{document}